\documentclass{sigchi}

\usepackage{cite}
\usepackage{balance}       % to better equalize the last page
\usepackage{graphics}      % for EPS, load graphicx instead 
\usepackage[T1]{fontenc}   % for umlauts and other diaeresis
\usepackage{txfonts}
\usepackage{mathptmx}
\usepackage[pdflang={en-US},pdftex]{hyperref}
\usepackage{color}
\usepackage{booktabs}
\usepackage{textcomp}
\usepackage{subcaption}
\usepackage{graphicx}% http://ctan.org/pkg/graphicx
\usepackage{amsfonts}
\usepackage{booktabs} % For formal tables
\usepackage{amsfonts}
\usepackage{amssymb}
\usepackage{xspace}
\usepackage{epstopdf}
\usepackage{enumitem}
\usepackage{algorithm}
\usepackage{algorithmic}
\usepackage{multirow}
\usepackage{booktabs}
\usepackage{array}
\usepackage{wrapfig}
\usepackage{pifont}
\usepackage{pbox}
\usepackage{multirow}
\usepackage{microtype}        % Improved Tracking and Kerning
\usepackage{ccicons}          % Cite your images correctly!
\usepackage{todonotes}
\usepackage{array}
\usepackage{tabulary}
\usepackage{xcolor}
\usepackage{amsmath}

\def\eg{\textit{e.g.}}

\def\nback{\textit{n}-back}
\usepackage{array}
\newcolumntype{L}[1]{>{\raggedright\let\newline\\\arraybackslash\hspace{0pt}}m{#1}}
\newcolumntype{C}[1]{>{\centering\let\newline\\\arraybackslash\hspace{0pt}}m{#1}}
\newcolumntype{R}[1]{>{\raggedleft\let\newline\\\arraybackslash\hspace{0pt}}m{#1}}
\newcolumntype{K}[1]{>{\centering\arraybackslash}p{#1}}
\newcommand*{\Comb}[2]{{}^{#1}C_{#2}}

\def\plaintitle{SIGCHI Conference Proceedings Format}

\def\emptyauthor{}
\def\plainkeywords{Authors' choice; of terms; separated; by
  semicolons; include commas, within terms only; required.}

\makeatletter
\def\url@leostyle{%
  \@ifundefined{selectfont}{
    \def\UrlFont{\sf}
  }{
    \def\UrlFont{\small\bf\ttfamily}
  }}
\makeatother
\def\pprw{8.5in}
\def\pprh{11in}

\definecolor{linkColor}{RGB}{6,125,233}
\hypersetup{%
  pdftitle={\plaintitle},
  pdfauthor={\emptyauthor},
  pdfkeywords={\plainkeywords},
  pdfdisplaydoctitle=true, % For Accessibility
  bookmarksnumbered,
  pdfstartview={FitH},
  colorlinks,
  citecolor=black,
  filecolor=black,
  linkcolor=black,
  urlcolor=linkColor,
  breaklinks=true,
  hypertexnames=false
}

\makeatletter
\def\@copyrightspace{\relax}
\makeatother

\begin{document}

\title{Investigating the generalizability of EEG-based Cognitive Load Estimation Across Visualizations}

\numberofauthors{6}
\author{%
  \alignauthor{%
    \textbf{Viral Parekh*}\\
    \affaddr{CVIT, IIIT Hyderabad} \\
    \affaddr{India} \\
    \email{viral@live.in} }
    \alignauthor{%
    \textbf{Maneesh Bilalpur*}\\
    \affaddr{CVIT, IIIT Hyderabad}\\
    \affaddr{India}\\
    \email{mbilalpur@gmail.com} }
    \alignauthor{%
    \textbf{Shravan Kumar}\\
    \affaddr{CVIT, IIIT Hyderabad}\\
    \affaddr{India}\\
    \email{shravankumar147@gmail.com} } 
    \vfil 
    \alignauthor{%
    \textbf{Stefan Winkler}\\
    \affaddr{Advanced Digital Sciences Center, UIUC}\\
    \affaddr{Singapore}\\
    \email{Stefan.Winkler@adsc-create.edu.sg} }
    \vfil 
    \alignauthor{%
    \textbf{C.V.Jawahar}\\
    \affaddr{CVIT, IIIT Hyderabad} \\
    \affaddr{India} \\
    \email{jawahar@iiit.ac.in} }
    \alignauthor{
    \textbf{Ramanathan Subramanian}\\
    \affaddr{Advanced Digital Sciences Center, UIUC}\\
    \affaddr{Singapore}\\
    \email{ramanathan.subramanian@ieee.org} } %
    }

\maketitle

\begin{abstract}
We examine if EEG-based cognitive load (CL) estimation is generalizable across the \textit{character}, \textit{spatial pattern}, \textit{bar graph} and \textit{pie chart}-based visualizations for the \nback~task. CL is estimated via two recent approaches: (a) Deep convolutional neural network~\cite{bashivan2015learning}, and (b) Proximal support vector machines~\cite{nback}. Experiments reveal that CL estimation suffers across visualizations motivating the need for effective {{machine learning}} techniques to benchmark visual interface usability for a given analytic task.
\end{abstract}



\section{Introduction}
\textit{A picture is worth a thousand words}-- this aphorism reflects the importance of \textit{\textbf{Information Visualization}} (InfoViz), which augments human analytical reasoning to solve complex real-world problems~\cite{Kohlhammer2011}. A key requirement of visual interfaces is that they should augment human perception while minimizing \textit{cognitive/mental workload}, which denotes the amount of mental resources expended during task performance. Cognitive load can be categorized as either natural or extraneous~\cite{chandler1991cognition}. In the InfoViz context, natural cognitive load is inherently imposed by the task on hand, while extraneous cognitive load depends on the visualization. Since visual interfaces are required to support exploratory analytics and provide insights, the use of usability heuristics or design questionnaires is unsuitable for evaluating interactive visualization interfaces (or Viz UIs)~\cite{Riche_beyondsystem,peck2013using}. 

\begin{figure}[htbp]
\includegraphics[width=0.20\linewidth]{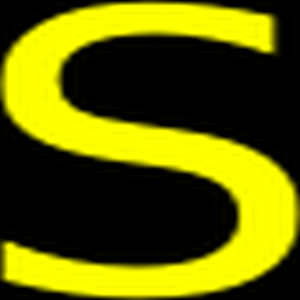}\hspace{0.1in}\includegraphics[width=0.20\linewidth]{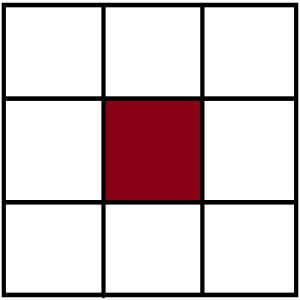}\hspace{0.1in}\includegraphics[width=0.26\linewidth]{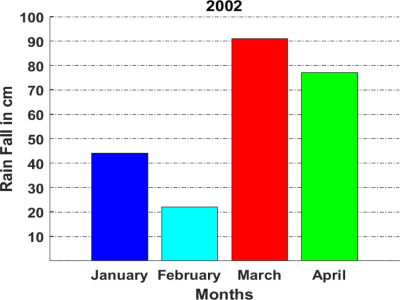}\hspace{0.02in}\includegraphics[width=0.26\linewidth]{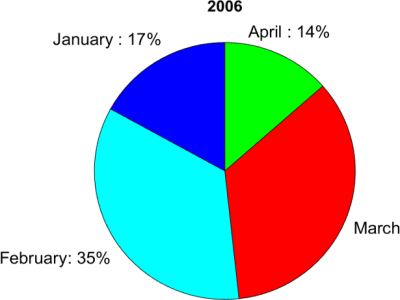}
\vspace{-.1cm}
\caption{\label{Exemplars} \textbf{Problem Statement:} Under varying mental workload levels induced by the \nback~task, we examined if there was any similarity in user cognitive behavior captured via EEG across four visualizations. Figure shows (from left to right) exemplar \textit{character}, \textit{position}, \textit{bar} and \textit{pie} visualizations.}
\vspace{-.2cm}
\end{figure}

{Neuroergonomics}, which examines human factors via neuroscientific methods, presents a viable alternative for evaluating Viz UIs. Lately, Viz evaluation via \textit{cognitive sensing} has been achieved by studying eye movement~\cite{Huang07,raschke2014visual,Korbach2018}, EEG~\cite{anderson2011user,nback,Bilalpur18} or fNIRS~\cite{peck2013using} activity patterns with light-weight, wireless devices~\cite{nback}. While being able to reliably measure cognitive load, these methods are nevertheless task plus visualization specific and non-generalizable~\cite{Ke2014}. This work examines if a single EEG framework can effectively assess (extraneous) memory workload across multiple visual interfaces under similar task difficulty. If cognitive load estimation (CLE) is generalizable, it would naturally enable Viz UI evaluation from neural data. Alternatively, a smart Viz UI could improve its current visualization with a potentially more intuitive one upon detecting high user workload.   

CLE generalizability has brightened with the success of deep convolutional neural networks (deep CNNs), which robustly learn problem-specific features and adapt effectively with minimal additional training~\cite{Shukla2017acm}. We examined if user EEG responses obtained for the \textit{character}, \textit{spatial pattern}, \textit{bar graph} and \textit{pie chart} visualizations under different mental workload levels induced by the \nback~task~\cite{peck2013using,bashivan2015learning,Ke2014,nback} had any similarities (Figure~\ref{Exemplars}). In lieu of learning a unified CLE model, we learned Viz-specific CLE models which were evaluated on (EEG data compiled for) other Viz types. We employed two state-of-the-art algorithms: deep CNN~\cite{bashivan2015learning}, and proximal SVM (pSVM)~\cite{nback}. Experiments suggest that (a) both models perform well when the train and test data are of the same Viz type, with pSVM outperforming deep CNN; and (b) the deep CNN achieves better CLE across Viz types, even if CLE performance deteriorates in such conditions. Our contributions are outlined on the left.

% \begin{figure*}[!h]
% \centering
% \includegraphics[width=.48\linewidth,height=3.2cm]{images/exp_protocol}
% \caption{Protocol timeline with 1-back exemplars.}
% \label{fig:easy}  
% \vspace{-.2cm}
% \end{figure*}

\begin{figure}[!h]
\centering
\includegraphics[width=.7\linewidth]{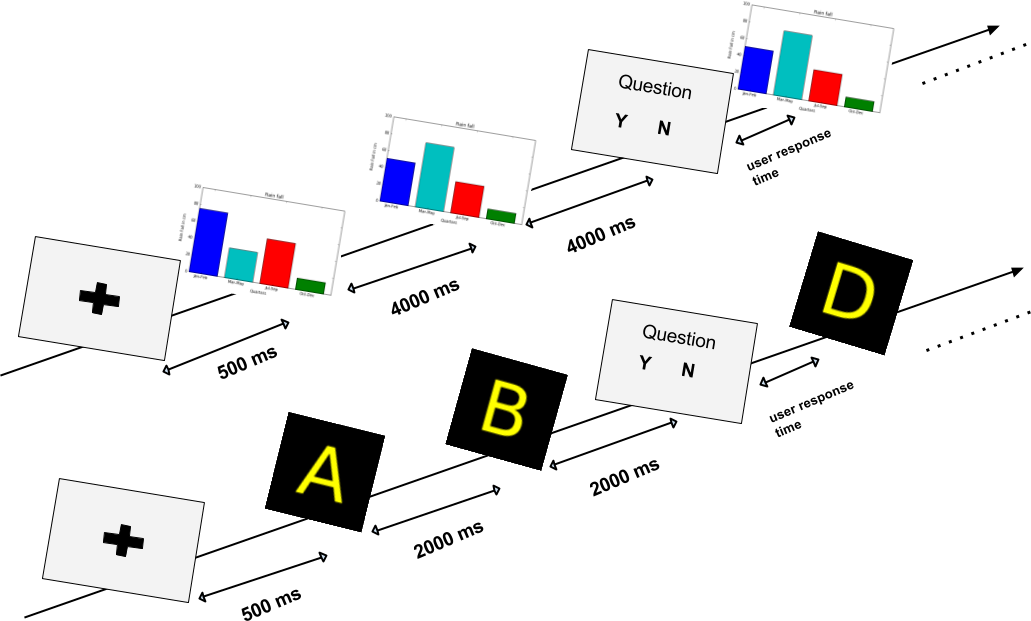}
\vspace{-0.1cm}
\caption{Protocol timeline with 1-back exemplars.}
\label{fig:easy}
\vspace{-0.2cm}
\end{figure}

\textbf{Contributions:} As a first step towards examining CLE generalizability, we examined \textit{{if there were similarities among the cognitive processes elicited by four different visualizations during \nback, based on EEG signals captured by a wireless headset}}. Wireless sensors suffer from low signal fidelity, but are more ecologically valid with respect to lab devices as they are convenient to use, enabling a non-intrusive and naturalistic user experience. 
\label{Sec:Contribution}

\subsubsection{Hypotheses}
Based on the experimental design, our hypotheses were as follows:

\begin{itemize}
\item[1.]  \textit{\textbf{{N-back} is more challenging with {bar} and pie:}} This is because users had to infer the measure of interest in the two slides via spatial and arithmetic inference before comparing for \textit{bar} and \textit{pie}, while \textit{char} and \textit{pos} required only a symbol/spatial pattern comparison. We posited that the challenge in \nback~for bar and pie would reflect via response times and accuracies observed for the four Viz types.        
\item[2.] \textit{\textbf{User performance will decrease for higher \nback:}} In spite of \textit{char} and \textit{pos} comparisons being easier than \textit{bar} and \textit{pie}, we nevertheless expected that (a) user performance would decrease for all Viz types with higher cognitive load (2 and 3-back), and (b) this should reflect via cognitive sensing such that EEG-based categorization of low/high mental workload should be facile irrespective if the Viz type.
\item[3.] \textit{\textbf{Cognitive processes for the char-pos and bar-pie Viz pairs should be similar:}} Following Hypothesis 1, even if the cognitive processes corresponding to the four Viz types are dissimilar, we still expected some compatibility between the CLE models for \textit{char} and \textit{pos}, and those for \textit{pie} and \textit{bar} given task similarity. 
\end{itemize}

\section{Materials and Methods}
\textbf{Stimuli, users and protocol:} We employed the four Viz types shown in Fig.\ref{Exemplars} in our study. These Viz types have been used previously~\cite{bashivan2015learning,peck2013using,nback}, and an \nback~task on these Viz types is designed to utilize the visual sensory pathway and working memory. Each \textit{char} stimulus comprised one of sixteen characters (selected randomly) centered on the screen, while each \textit{pos} stimulus was a $3\times3$ spatial grid with one of nine blocks highlighted. The \textit{bar} and \textit{pie} stimuli were generated from real-life rainfall data from January to April. Bar graphs depicted raw rainfall levels (marked in cm), while pie charts encoded these values as percentages. 20 graduate students (11 male, age 24.4$\pm$2.1) with normal or corrected vision took part in our study, which was approved by the local ethics committee.  

The standard \nback~design (Fig.\ref{fig:easy}) was used over 24 blocks constituting a user session. Within each block, users were presented with a series of slides, and needed to compare the current slide $s$ with the $s-n^{th}$ slide to make a \textit{yes} or \textit{no} decision where $n$ ranged between 0--3. 0-back required comparison with a pre-defined value/pattern. For \textit{char} and \textit{pos}, we asked if a letter/position matched with $n$ slides before. For \textit{pie} and \textit{bar}, we asked whether a specific measurement (\eg, rainfall in March) was greater than $n$ slides ago. As bar and pie charts are designed to use human visuospatial ability, values to compare had to be inferred by users from the \textit{pie} and \textit{bar} graphs. Identical colors were used to encode the \textit{bar} and \textit{pie} charts. Users recorded responses via a radio button and to minimize fatigue, each user session was split into two 30 minute halves. Each block contained 12 slides from one Viz type and each user session comprised 288 presentations (2 halves/session $\times$ 12 blocks/half $\times$ 12 slides/block).

\begin{table}[!ht]
\centering
\begin{tabular}{|cccc|}
\hline	
\textbf{Response}& &\multicolumn{1}{p{0.25\marginparwidth}}{\centering{\textbf{RT}}} & \multicolumn{1}{p{0.25\marginparwidth}|}{\centering{\textbf{RA}}} \\ \toprule \bottomrule
\textbf{Predictor}&\textbf{df}&\multicolumn{1}{p{0.25\marginparwidth}}{\centering{\textbf{F}}}&\multicolumn{1}{p{0.25\marginparwidth}|}{\centering{\textbf{F}}} \\
\textbf{Viz Type}&3&3.96*&74.04* \\
\textbf{N-back}&3&27.17*&24.75* \\
\textbf{Interaction}&9&4.33*&1.62 \\
\textbf{Error}&304& & \\
\textbf{Total}&316& & \\ \hline
	
\end{tabular}%\vspace{-.2cm}
\caption{ANOVA summary for RTs and RAs. df denotes degrees of freedom and * denotes significance at $p<0.01$.}\label{tab:beh_stat}
	\vspace{-.2cm}
\end{table}

The timeline within each block is as shown in Fig.\ref{fig:easy}. Following a 500 ms fixation cross, the display duration of each successive slide was set to 2s for \textit{char}/\textit{pos}, and 4s for \textit{bar}/\textit{pie} slides. Longer display times for \textit{bar} and \textit{pie} were set as users needed to infer measures of interest via spatial and arithmetic means, unlike a mere symbol/pattern comparison for \textit{char} and \textit{pos}. Users had to record their responses within a 10s limit and each instance involving a user response is denoted as a \textit{trial}. Of the 24 blocks, 12 were designed to be 0 or 1-back, while the other 12 were~2 and 3-back. The order of Viz types was randomized and the number of 0/1/2/3-back blocks remained identical across users. Our study employed a 4$\times$4 within-subject design involving two factors-- the \nback~level (0,1,2,3 back) and the Viz type (\textit{char}, \textit{pos}, \textit{bar}, \textit{pie}). Users' neural activity was recorded via the 14-channel \textit{Emotiv} wireless EEG device as they performed the experiment (no subjective impressions were collected). 

%%%%%%%%%%%%%%%%%%%%%%%%%%%%%%%%%%%%%%%%%%%%%%%%%%%%(a) 

%According to theories like Cognitive Load Theory and Cognitive Thoery of Multimedia Learning, the optimal learning conditions are characterized by providing challenges to learners without cognitive overload\cite{gerjets2014cognitive}. In our task, we have made sure that the n-back task while presenting challenges, had no components which the participants felt extremely hindering their ability to perform the task.

\begin{figure}[!h]
\centering
\includegraphics[width=.48\linewidth,height=3.2cm]{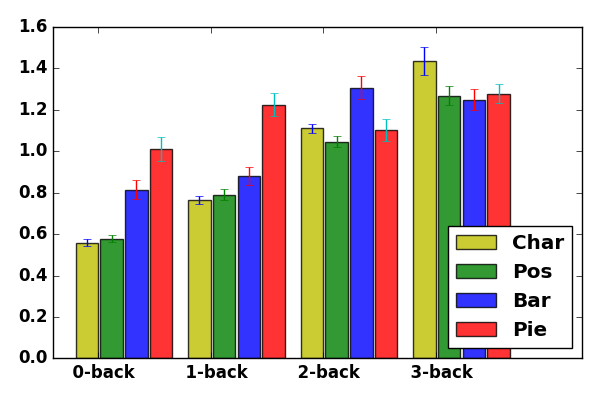}\hspace{0.1cm}\includegraphics[width=.48\linewidth,height=3.2cm]{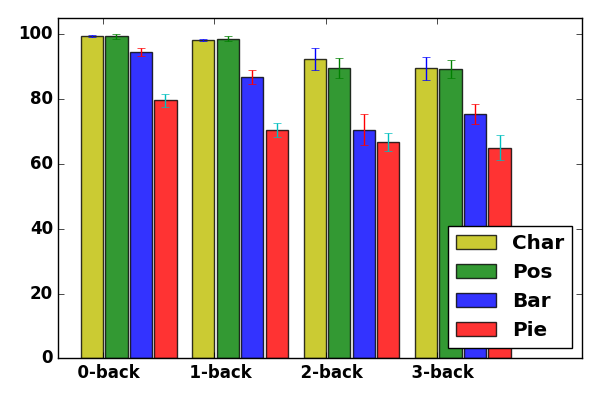}
\vspace{-0.1cm}
\caption{(left) RTs (in seconds) and (right) RAs for different Viz and \nback~types. Error bars denote unit standard error.}
\label{fig:rts_rrs}
\vspace{-0.2cm}
\end{figure}

Our hypotheses and ANOVA analyses of user responses in the form of response times (RTs) and response accuracies (RAs) are presented on the left.  Supporting H1, user responses were much faster for \textit{char} and \textit{pos} in 0 and 1-back, while RTs for all four Viz types become very comparable from 2-back onwards (Fig.\ref{fig:rts_rrs}). Increasing RT with \nback~type reveals that task difficulty increases with $n$ due to greater load on working memory. A two-way ANOVA on RTs revealed the main effect of Viz type, \nback~type and their interaction effect (Table~\ref{tab:beh_stat}). Consistent with H2(a), there was a steady decline in user performance with increasing \nback. Close-to-ceiling performance was noted with \textit{pos} and \textit{char} for 0 and 1-back, whereas less than 80\% accuracy was noted for \textit{pie} even for 0-back. RAs considerably decreased across Viz types for 2 and 3-back, and an ANOVA on RAs revealed the main effect of Viz type and \nback~level (Table~\ref{tab:beh_stat}).

\textit{\textbf{User behavioral data clearly validate Hypotheses 1 and 2(a).}} The challenge posed by \nback~with \textit{bar} and \textit{pie} is reflected via higher response times, and sharply lower accuracies for these Viz types (Fig.\ref{fig:rts_rrs}). \textit{Char} and \textit{pos} visualizations appear comparable and result in very similar RTs and RAs. However, bar graphs seem to be more easy to interpret (reflected via significantly higher RAs) than pie charts even as corresponding RTs only differ slightly. Our results agree with~\cite{Cleveland.McGill1984} which poses an estimation task, but differ from ~\cite{peck2013using} which poses a comparison task to find that \textit{bar}-\textit{pie} performance differences are user-specific than holistic.

\section{EEG-based CLE: Results and Discussion}
We recorded EEG data via the wireless 14-channel \textit{Emotiv EPOC} headset. EEG data is contaminated by various noise sources (power-line noise, muscle and eye movement based artifacts). Upon (i) removing corrupt EEG recordings, (ii) band-pass filtering the EEG signal to within 0.1--45 Hz and (iii) visually removing noisy signal components post Independent Component Analysis (ICA) to remove muscular, head and eye movement artifacts, we extracted \textit{two second epochs} from the period immediately preceding user response from each trial. We then estimated CLE with EEG data employing the deep CNN~\cite{bashivan2015learning} and proximal SVM~\cite{nback} based algorithms. The key difference between the deep CNN and pSVM methods is that topography maps preserving spatial and spectral EEG structure are input to the deep CNN (with structure similar to the VGG architecture~\cite{simonyan2014very}) in~\cite{bashivan2015learning}, while pSVM~\cite{nback} learns from a vectorized, maximum relevance and minimum redundancy feature set identified via an information theoretic approach.

\begin{table}[!ht]
\centering
\begin{tabular}{|c|ccccc|}
\hline
	\textbf{Type} & \textbf{0-back} & \textbf{1-back} & \textbf{2-back} & \textbf{3-back} & \textbf{Total} \\ \hline
	\textit{Char} & 356 & 199 & 364 & 167 & 1086 \\ 
	\textit{Pos} & 377 & 175 & 165 & 376 & 1093 \\ 
	\textit{Bar} & 203 & 391 & 194 & 390 & 1178 \\ 
	\textit{Pie} & 193 & 401 & 393 & 192 & 1179 \\ 
	\textbf{Total} & 1129 & 1166 & 1116 & 1125 & \textbf{4536} \\ \hline
\end{tabular}
\vspace{-.1cm}
\caption{\label{tab:expts} EEG epoch distribution based on Viz and \nback~type.}
\vspace{-.4cm}
\end{table}
We performed six $(\Comb{4}{2})$ disjoint pairwise classifications (\eg, 0-back vs 1-back). To examine similarities among cognitive processes induced by different Viz types, we trained models with labeled epochs for one Viz type, and tested them with epochs for another Viz type. Classification was performed in a user-independent setting, and given the varying epoch counts for each \nback~condition (due to \nback~design and noise removal (Table~\ref{tab:expts})), we employed F1-score as our performance metric. Table~\ref{results:dnn} presents mean F1 scores achieved with 10-fold cross validation. Within-Viz results are denoted in blue along the table diagonal and the highest F1 score obtained across  \nback~categorizations for a given Viz pair is denoted in bold. \textbf{\textit{For both algorithms, the best F1-score in 13 of the 16 conditions corresponds to \textit{low} vs high cognitive load categorization, validating Hypothesis 2(b)}} and implying that coarse-grained benchmarking of mental workload is more feasible than fine-grained differentiation. Comparing the maximum within-Viz F1-scores, we find that pSVM outperforms deep CNN. Also, cross-Viz \nback~categorization is inferior to within-Viz, particularly for pSVM. This suggests (a) possible EEG signal differences, and consequently EEG features obtained for the four Viz types, and (b) the pSVM method which performs vector classification is unable to effectively deal with these differences. Cross-viz results achieved with structure-preserving deep CNN are relatively robust. Also, observed results \textit{\textbf{only provide limited support to H3 that the cognitive processes for the char-pos and bar-pie pairs may be similar}}. Overall, {these results call for more research towards generalized CLE benchmarking, specifically motivating the need to encode brain activations more efficiently and robustly, even while conveying that available tools hold some promise in this direction}.

\begin{table}[!ht]
\fontsize{7}{7}\selectfont
\renewcommand{\arraystretch}{1.2}
%\flushleft
\begin{tabular}{|K{0.75cm}|K{0.75cm}|K{0.75cm}K{0.75cm}|K{0.75cm}K{0.75cm}|K{0.75cm}K{0.75cm}|K{0.75cm}K{0.75cm}|}
\hline
\multicolumn{2}{|l|}{} & \multicolumn{2}{c|}{\textbf{char}} & \multicolumn{2}{c|}{\textbf{pos}} & \multicolumn{2}{c|}{\textbf{bar}} & \multicolumn{2}{c|}{\textbf{pie}} \\ %\cline{3-10} 
\multicolumn{2}{|l|}{\multirow{-2}{*}{}} & \textbf{pSVM} & \textbf{CNN} & {\textbf{pSVM}} & \textbf{CNN} & {\textbf{pSVM}} & \textbf{CNN} & {\textbf{pSVM}} & \textbf{CNN} \\ \hline
 & \textbf{0 vs 1} & {\textcolor{blue}{0.72}} & {\textcolor{blue}{0.63}} & \multicolumn{1}{c}{0.54} & 0.68 & 0.27 & 0.52 & 0.47 & 0.61 \\
 & \textbf{0 vs 2} & {\textcolor{blue}{0.65}} & {\textcolor{blue}{0.66}} & \multicolumn{1}{c}{0.56} & 0.73 & 0.51 & \textbf{0.71} & 0.52 & 0.59 \\
 & \textbf{0 vs 3} & {\textcolor{blue}{0.83}} & {\textcolor{blue}{0.69}} & \multicolumn{1}{c}{0.48} & 0.68 & 0.48 & 0.54 & 0.51 & 0.63 \\
 & \textbf{1 vs 2} & {\textcolor{blue}{0.82}} & {\textcolor{blue}{0.71}} & \multicolumn{1}{c}{\textbf{0.71}} & \textbf{0.74} & 0.38 & 0.54 & 0.45 & 0.59 \\
 & \textbf{1 vs 3} & {\textcolor{blue}{0.77}} & {\textcolor{blue}{0.64}} & \multicolumn{1}{c}{0.44} & 0.49 & \textbf{0.62} & 0.62 & \textbf{0.64} & \textbf{0.68} \\
\multirow{-6}{*}{\textbf{char}} & \textbf{2 vs 3} & {\textcolor{blue}{\textbf{0.86}}} & {\textcolor{blue}{\textbf{0.80}}} & \multicolumn{1}{c}{0.46} & 0.58 & 0.25 & 0.35 & 0.57 & 0.63 \\ \hline
 & \textbf{0 vs 1} & 0.53 & \textbf{0.68} & \multicolumn{1}{c}{{\textcolor{blue}{0.76}}} & {\textcolor{blue}{0.64}} & 0.53 & 0.52 & 0.28 & 0.44 \\
 & \textbf{0 vs 2} & 0.59 & 0.65 & \multicolumn{1}{c}{{\textcolor{blue}{0.74}}} & {\textcolor{blue}{\textbf{0.72}}} & \textbf{{0.65}} & \textbf{0.69} & 0.47 & 0.48 \\
 & \textbf{0 vs 3} & 0.47 & 0.62 & \multicolumn{1}{c}{{\textcolor{blue}{0.70}}} & {\textcolor{blue}{0.60}} & 0.44 & \textbf{0.69} & 0.42 & \textbf{0.70} \\
 & \textbf{1 vs 2} & \textbf{0.66} & 0.65 & \multicolumn{1}{c}{{\textcolor{blue}{\textbf{0.80}}}} & {\textcolor{blue}{0.64}} & 0.57 & 0.61 & \textbf{0.50} & 0.54 \\
 & \textbf{1 vs 3} & 0.50 & 0.58 & \multicolumn{1}{c}{{\textcolor{blue}{0.77}}} & {\textcolor{blue}{0.68}} & 0.49 & 0.60 & 0.40 & 0.53 \\
\multirow{-6}{*}{\textbf{pos}} & \textbf{2 vs 3} & 0.49 & 0.52 & \multicolumn{1}{c}{{\textcolor{blue}{0.79}}} & {\textcolor{blue}{0.45}} & 0.60 & 0.67 & 0.34 & 0.47 \\ \hline
 & \textbf{0 vs 1} & 0.29 & 0.53 & \multicolumn{1}{c}{\textbf{0.60}} & 0.63 & {\textcolor{blue}{0.70}} & {\textcolor{blue}{0.63}} & 0.47 & 0.66 \\
 & \textbf{0 vs 2} & 0.52 & 0.58 & \multicolumn{1}{c}{0.59} & \textbf{0.75} & {\textcolor{blue}{0.64}} & {\textcolor{blue}{0.61}} & \textbf{{0.59}} & 0.48 \\
 & \textbf{0 vs 3} & 0.53 & 0.58 & \multicolumn{1}{c}{0.53} & 0.74 & {\textcolor{blue}{\textbf{0.81}}} & {\textcolor{blue}{\textbf{0.83}}} & {0.51} & \textbf{0.77} \\
& \textbf{1 vs 2} & 0.39 & 0.41 & \multicolumn{1}{c}{0.57} & 0.46 & {\textcolor{blue}{0.72}} & {\textcolor{blue}{0.69}} & 0.50 & 0.53 \\
& \textbf{1 vs 3} & \textbf{0.58} & \textbf{0.64} & \multicolumn{1}{c}{0.55} & 0.54 & {\textcolor{blue}{0.67}} & {\textcolor{blue}{0.78}} & 0.44 & 0.70 \\
\multirow{-6}{*}{\textbf{bar}} & \textbf{2 vs 3} & 0.26 & 0.37 & \multicolumn{1}{c}{0.40} & 0.65 & {\textcolor{blue}{0.72}} & {\textcolor{blue}{0.65}} & 0.51 & 0.63 \\ \hline
 & \textbf{0 vs 1} & 0.45 & 0.48 & \multicolumn{1}{c}{0.23} & 0.52 & 0.50 & \textbf{0.66} & {\textcolor{blue}{0.73}} & {\textcolor{blue}{0.54}} \\
 & \textbf{0 vs 2} & 0.56 & \textbf{0.63} & \multicolumn{1}{c}{0.41} & 0.62 & 0.41 & 0.62 & {\textcolor{blue}{0.77}} & {\textcolor{blue}{0.58}} \\
& \textbf{0 vs 3} & \textbf{0.62} & 0.61 & \multicolumn{1}{c}{0.57} & \textbf{0.67} & \textbf{0.59} & 0.58 & {\textcolor{blue}{0.70}} & {\textcolor{blue}{0.68}} \\
 & \textbf{1 vs 2} & {0.50} & 0.56 & \multicolumn{1}{c}{0.50} & 0.62 & 0.43 & 0.62 & {\textcolor{blue}{0.62}} & {\textcolor{blue}{0.59}} \\
& \textbf{1 vs 3} & 0.61 & 0.56 & \multicolumn{1}{c}{0.46} & 0.39 & 0.57 & 0.57 & {\textcolor{blue}{0.65}} & {\textcolor{blue}{\textbf{0.74}}} \\

\multirow{-6}{*}{\textbf{pie}} & \textbf{2 vs 3} & 0.58 & 0.60 & \multicolumn{1}{c}{0.28} & 0.43 & 0.49 & 0.39 & {\textcolor{blue}{\textbf{0.78}}} & {\textcolor{blue}{0.68}} \\ 
\hline
\end{tabular} 
\vspace{-.1in}
%\small
\caption{ Cross-Viz CLE with pSVM~\protect\cite{nback} and deep CNN~\protect\cite{bashivan2015learning}. Training and test Viz type denoted by rows and columns. }\label{results:dnn} \vspace{-.2in}
\end{table}

\balance{}

% REFERENCES FORMAT
% References must be the same font size as other body text.
% \bibliographystyle{SIGCHI-Reference-Format}
\bibliography{output}

\end{document}